\documentclass[unnumsec,webpdf,contemporary,large]{brsl-template}%

\usepackage{microtype}
\usepackage{cleveref}

\usepackage{paratype} 
\usepackage[T1]{fontenc}

\usepackage{color,soul}

\usepackage{xcolor}
\definecolor{light-gray}{gray}{0.95}
\newcommand{\code}[1]{\colorbox{light-gray}{\texttt{#1}}}

\graphicspath{{Fig/}}

\theoremstyle{thmstyleone}%
\theoremstyle{thmstyletwo}%
\theoremstyle{thmstylethree}%

\begin{document}

\journaltitle{AI Frontiers Initiative}
\copyrightyear{2024}
\pubyear{November 2024}
\appnotes{\hspace{1mm}}

\firstpage{1}


\title[Whack-a-Chip]{Whack-a-Chip: \Large{The Futility of Hardware-Centric Export Controls}}

\author[1,2,$\ast$]{Ritwik Gupta\ORCID{0000-0001-7608-3832}}
\author[1]{Leah Walker}
\author[1]{Andrew W. Reddie\ORCID{0000-0003-3231-8307}}

\authormark{Ritwik Gupta et al.}

\address[1]{\orgdiv{Berkeley Risk and Security Lab}, \orgname{University of California, Berkeley}}
\address[2]{\orgdiv{Berkeley AI Research Lab}, \orgname{University of California, Berkeley}}

\corresp[$\ast$]{Corresponding author: \href{ritwikgupta@berkeley.edu}{ritwikgupta@berkeley.edu}}



\abstract{
U.S. export controls on semiconductors are widely known to be permeable, with the People's Republic of China (PRC) steadily creating state-of-the-art artificial intelligence (AI) models with exfiltrated chips. This paper presents the first concrete, public evidence of how leading PRC AI labs evade and circumvent U.S. export controls. We examine how Chinese companies, notably Tencent, are not only using chips that are restricted under U.S. export controls but are also finding ways to circumvent these regulations by using software and modeling techniques that maximize less capable hardware. Specifically, we argue that Tencent's ability to power its Hunyuan-Large model with non-export controlled NVIDIA H20s exemplifies broader gains in efficiency in machine learning that have eroded the moat that the United States initially built via its existing export controls. Finally, we examine the implications of this finding for the future of the United States' export control strategy.
}

\maketitle

\section{Introduction}

In an effort to maintain a technical edge, the United States has sought to limit the People’s Republic of China’s (PRC) artificial intelligence (AI) research and development by pursuing export controls on hardware (notably chips). As part of this strategy, the United States has created a series of limits on AI hardware procurement including not only chips but also the manufacturing equipment needed to create them~\cite{bensonUpdatedOctober72023}. This strategy reflects the assumption that limiting a country’s access to cutting-edge computing hardware limits the ability to create cutting-edge ``AI'' and other advancements. In reality, we argue that both the strategy and its underlying assumption have proven unreliable.

First, major Chinese AI firms---despite the export controls that they face---are still managing to access high-end export-controlled chips. Second, Chinese AI labs have leveraged advancements in machine learning (ML) training tools to successfully train state-of-the-art (SOTA) models on lower quality, non-export controlled chips (including NVIDIA’s H20 GPUs), demonstrating that an export strategy based on hardware thresholds can be overcome through better software. The latter reality reflects an evolution in the field of machine learning at large and illustrates the importance of keeping abreast of the cutting edge of machine learning and compute efficiency as part of forecasting competitor capability gains.

The strategy of overcoming limited hardware resources with better software is not unique to the PRC---American academic labs are bellwethers for this phenomenon. Both are resorting to clever optimizations~\cite{khandelwal100K100Days2024} due to their inability to source high-quality chips against better-resourced organizations. Interestingly, these trends in academic computing can also offer a useful proxy to understand ways in which the PRC might achieve its AI goals in spite of an unreliable stream of advanced hardware.

In light of these challenges, this paper examines recent AI models released by Chinese companies and examines evidence of Chinese firms using export-controlled chips and cases of efficiency gains with non-export-controlled chips. Specifically, this paper focuses on case studies and examples from Tencent, the Chinese video game juggernaut and Chinese leader in AI research and development. This work represents the first public analysis demonstrating how Chinese companies are successfully using NVIDIA’s H20 GPUs, explicitly allowed under U.S. export controls, to train state-of-the-art AI, including Tencent’s recently released Hunyuan-Large Model. We highlight Tencent's previous use of cutting-edge, export-controlled chips like the NVIDIA A100 for its research and show how codebase signals can reveal which chips are used for training.

While there has been previous reporting on the porous nature of the U.S. export regime~\cite{yeFocusChinasUnderground2023, linWSJNewsExclusive2023, olcottChineseAIGroups2023}, this paper takes a deeper dive, focusing on the methods and techniques enabling Chinese firms to overcome its reliance on export-controlled chips and the acquisition and use of restricted microelectronics. Finally, we analyze the broader implications of Beijing’s evolving evasion strategies. 

\section{Chinese Firms are Outpacing Hardware Restrictions}
On November 4, 2024, the Chinese tech giant Tencent released ``Hunyuan-Large,'' the largest, open-source, transformer-based mixture-of-experts model which achieves state-of-the-art performance on multiple downstream tasks~\cite{sunHunyuanLargeOpenSourceMoE2024}.

Despite its state-of-the-art nature and size, a close examination of Hunyuan-Large reveals that it was not trained on state-of-the-art hardware. Rather Hunyuan-Large was trained on an export control-compliant chip, the NVIDIA H20, a feat made possible by a series of techniques used to maximize the value of throttled hardware. 

\subsection{Hunyuan-Large: A SOTA, Open-Source LLM}
Hunyuan-Large represents Tencent’s newest, open-source LLM. When compared to other state-of-the-art LLMs such as Meta’s Llama 3.1~\cite{llamateamatmetaLlama3Herd2024}, Mixtral-8x22B~\cite{aiCheaperBetterFaster2024}, and DeepSeek-V2~\cite{deepseekaiDeepSeekV2StrongEconomical2024}, it achieves state-of-the-art performance on multiple downstream benchmarks such as MMLU~\cite{hendrycksMeasuringMassiveMultitask2021} and CommonsenseQA~\cite{talmorCommonsenseQAQuestionAnswering2019}.

On November 5, 2024, Hunyuan-Large’s training codebase\footnote{\href{https://github.com/RitwikGupta/Tencent-Hunyuan-Large}{https://github.com/RitwikGupta/Tencent-Hunyuan-Large}. We link to a forked version of the repository to ensure that it is frozen in place, as analyzed.} and weights\footnote{\href{https://huggingface.co/tencent/Tencent-Hunyuan-Large}{https://huggingface.co/tencent/Tencent-Hunyuan-Large}} were openly released by Tencent on GitHub and Hugging Face, respectively. These websites are industry standards for sharing open-source code and model weights. In addition to releasing weights on Hugging Face, Tencent created a Hugging Face Space to interact with the LLM. The project’s README explicitly states that the model was trained on NVIDIA’s H20 GPUs.

The NVIDIA H20 represents NVIDIA’s GPU offering that is compliant with U.S. export controls on advanced semiconductors; it offers only three-quarters of the performance of the comparable NVIDIA H100~\cite{mujtabaNVIDIAsChinaCompliantH202024}. However, despite its lower performance, the H20 boasts 96GB of VRAM compared to the H100’s common 80GB configuration.\footnote{The H100 is also available in 96GB and 144GB configurations, but they are less common.} In part driven by memory performance, the H20 has been in the spotlight as the next advanced computing chip to become the subject of export control~\cite{zuhairNVIDIAsH20AI2024}.

This whack-a-mole approach to limiting chips based on performance thresholds is deeply flawed and is blind to the intended downstream use of models trained on controlled chips. For example, metrics in the newly introduced export control classification number (ECCN 3A090), such as total processing performance, can be sidestepped through the implementation of increasingly efficient machine learning training toolkits. Software advancements are making old hardware increasingly useful.

\subsection{Efficiently Training SOTA Models on Nerfed Hardware with Better Software}
Models are becoming smaller, requiring less compute to train, and converging faster as machine learning engineering research unlocks efficiencies in old hardware~\cite{guptaDataCentricAIGovernance2024}. Here, we analyze the tools, techniques, and procedures utilized by projects such as Hunyuan-Large to train state-of-the-art models despite working with throttled hardware.

\subsubsection{Ensembles and Mixtures-of-Experts}
Model ensembling represents a common technique in which the outputs from multiple models are combined to address variance in underlying data distributions or weaknesses in a single, monolithic model with individual, small models ensembled together using less compute and providing higher accuracy than their larger counterparts~\cite{kondratyukWhenEnsemblingSmaller2020}. Mixture-of-experts (MoE) is a related technique that has seen a resurgence with modern LLMs. With a fixed compute budget, MoEs allow for the training of bigger models in aggregate while achieving the same accuracy as a monolithic model~\cite{sansevieroMixtureExpertsExplained2023}. Tencent leveraged this technique for Hunyuan-Large to train a strong model efficiently on comparatively limited hardware.

\subsubsection{Mixed-Precision Training via bfloat16}
Hunyuan-Large also uses\footnote{\href{https://huggingface.co/tencent/Tencent-Hunyuan-Large/blob/main/Hunyuan-A52B-Pretrain/config.json\#L46}{Link to Hugging Face}} mixed-precision training,\footnote{\href{https://github.com/RitwikGupta/Tencent-Hunyuan-Large/blob/main/train/ds_zero3_offload_no_auto.json\#L11}{Link to repository}} made possible by the use of the bfloat16 data type. The brain floating point number was introduced to the public by Google Brain in 2019 as a shortened, 16-bit version (bfloat16) of the standard 32-bit IEEE 754 single-precision floating-point format (float32). With a truncated significand such as with float16, but with the dynamic range of float32, the bfloat16 representation is memory efficient and can run calculations faster. This representation has become the de facto standard for both half- and mixed-precision training~\cite{micikeviciusMixedPrecisionTraining2018}.

Mixed precision training has been shown to train models up to 2.5$\times$ faster than full-precision training with float32 on advanced GPUs such as the NVIDIA A100~\cite{huangIntroducingNativePyTorch2020}. This training paradigm allows for larger models, larger batches, or larger inputs while achieving the same accuracy as full-precision training. NVIDIA introduced bfloat16 support with its Ampere micro-architecture (such as the A100) and is available in all following architectures, including Hopper (the family of chips that the H20 falls under).

\subsubsection{Quantization}
Using less precise number representations like float16, int16, or even binary comes with drastic memory savings. Larger models can run faster on constrained resources. Recent advances in quantization have reduced model sizes and increased throughput by 16$\times$ while maintaining, or only slightly degrading, model accuracy~\cite{gholamiSurveyQuantizationMethods2021}.

As an example, the throughput of an NVIDIA A100 with float32 operations on tensor cores is 156 TFLOPS, while it can achieve a throughput of int8 operations on tensor cores of 624 TFLOPS, an idealized 3$\times$ increase in throughput.\footnote{\href{https://www.nvidia.com/content/dam/en-zz/Solutions/Data-Center/a100/pdf/nvidia-a100-datasheet-us-nvidia-1758950-r4-web.pdf}{NVIDIA A100 Tensor Core GPU Datasheet}.} Meanwhile Meta's Llama-3-8B demonstrates only a $0.019\pm0.003\%$ reduction in perplexity when being quantized from float16 to int8~\cite{pochinkovComparingQuantizedPerformance2024}.

\subsubsection{Large VRAM and Sharded Training}
The most significant limiting factor to training large models is the amount of video random access memory (VRAM) available on GPUs. Larger models do not fit on individual GPUs. If multiple GPUs are available, then sharding (splitting) a model across these GPUs allows large models to be trained. This is no easy feat---effective training in a sharded, multi-node environment has been the subject of continual study in modern machine learning research~\cite{gholamiAIMemoryWall2024, wanEfficientLargeLanguage2024, menghaniEfficientDeepLearning2023}. Sharding comes with communication overhead, drastically reducing the speed of model training. Where possible, having high VRAM GPUs can be more important than having GPUs with a large amount of fast cores. In fact, many resource-constrained organizations such as academic research labs aim to maximize the VRAM/dollar ratio rather than aggregate core counts~\cite{khandelwal100K100Days2024}.

As mentioned above, the China-export-compliant NVIDIA H20 offers 96GB of VRAM. In a world where American scientists and companies struggle to get H100s~\cite{morganWhatWhenYou2023} with comparable amounts of VRAM, they resort to using older 40/80GB A100s instead. The Chinese market arguably has an edge over the American market when it comes to high VRAM chips as the default chip available to them has high VRAM.

Moreover, GPUs are scarce across the globe. To make efficient use of this limited resource, techniques and libraries including fully-sharded data parallelism~\cite{zhaoPyTorchFSDPExperiences2023} and Microsoft’s DeepSpeed~\cite{rasleyDeepSpeedSystemOptimizations2020} are used to train large models fast. These techniques shard model parameters, gradients, and even optimizer states across multiple GPUs which themselves are split across multiple nodes. The library does this in such a way that minimizes communication overhead and any ``bubbles,'' periods where hardware goes unused, in the parallel pipeline.

Constrained to using ``nerfed'' NVIDIA H20 GPUs, the Hunyuan-Large LLM is trained using Microsoft’s DeepSpeed library. Specifically, they utilize ZeRo Stage 3 training which partitions all model states (parameters, gradients, and optimizer states) across networked GPUs, CPUs, and RAM~\cite{rajbhandariZeROMemoryOptimizations2020}.

Other tricks to overcome the limitations of scarce GPU resources include gradient accumulation in which multiple batches of data are passed through the model before one step of optimization, effectively simulating larger batch sizes for optimization. This is a common technique when high VRAM GPUs are unavailable. Since Tencent has access to NVIDIA H20 GPUs, they did not use gradient accumulation to train Hunyuan-Large but other Chinese firms may for other models.

\subsubsection{Effectively Using Networked GPUs}
When training in a sharded setting, communication overhead drastically slows down the process. Traditionally, GPUs communicate with each other via the CPU, which can be slow. GPUDirect Remote Direct Memory Access (RDMA) solves this problem by allowing GPUs to share memory directly, drastically decreasing overhead. Rather than transfer data from GPU VRAM to CPU RAM and back to another GPU’s VRAM, GPUDirect RDMA allows GPUs to communicate with each other directly via the NVLink or PCIe network fabric locally, or InfiniBand or Ethernet across nodes.

The training script for Hunyuan-Large provided by the model developers explicitly enabled GPUDirect RDMA over InfiniBand (\code{NCCL\_P2P\_DISABLE=0}, \code{NCCL\_IB\_CUDA\_SUPPORT=1}, and \code{NCCL\_NET\_GDR\_LEVEL=2}) with eight GPUs per node. This training was likely conducted in a high-performance computing environment such as a data center. Eight, power-hungry GPUs generally fit in a 2U or 4U server rack, not in consumer PC cases. Furthermore, the training script bonds multiple InfiniBand network interfaces (\code{NCCL\_IB\_DISABLE=0} and \code{NCCL\_IB\_HCA=...}); InfiniBand and bonded network interfaces are commonly utilized in data centers. Importantly, limiting the ability of China to network GPUs together to build a “supercomputer” is the stated goal of U.S. export controls on semiconductors~\cite{bureauofindustryandsecurityImplementationAdditionalExport2022}.

\subsection{Overriding Software Limitations in Hardware}
Notably, NVIDIA limits GPUDirect RDMA to only its data center GPU offerings, excluding clusters built using consumer GPUs. Thus consumer GPUs, including the NVIDIA RTX 4090---common in academic labs---are left to suffer communication overhead that GPUDirect RDMA alleviates in data center GPUs.

The limiting of direct peer-to-peer communications is not done in hardware---it is ``soft locked'' through proprietary NVIDIA drivers. Tiny Corp., a startup developing AI ``supercomputers,'' reverse-engineered and publicly released custom drivers for the RTX 4090 which enabled peer-to-peer communications in June 2024---opening this soft lock.\footnote{\href{https://github.com/tinygrad/open-gpu-kernel-modules}{https://github.com/tinygrad/open-gpu-kernel-modules}} Reporting has shown that China is using RTX 3090 and 4090 GPUs for AI workloads~\cite{harperDesperateChineseFactories2023}; drivers such as the one released by Tiny Corp. could increase the effectiveness of these bootleg data centers.

\section{Chinese Firms Bypass Chip Restrictions}
\label{sec:bypass}

Export controls on advanced semiconductors such as graphics processing units (GPUs) were expanded by the United States in conjunction with the 2022 CHIPS and Science Act under the Biden-Harris Administration. Specifically, the newly introduced export control classification number (ECCN 3A090) controls the export of semiconductors marketed for use in data centers that meet a specific performance threshold to Chinese and Russian entities. As we have previously discussed, certain thresholds, such as inter-chip bandwidth, were not adequately designed to account for the types of models that could be trained on chips below these thresholds~\cite{guptaAcceleratingEvolutionAI2023}. These thresholds were subsequently revised to address potential ‘gray zone’ loopholes~\cite{bureauofindustryandsecurityCommerceStrengthensRestrictions}. In practice, this made export controls stricter and included large, single-wafer accelerators such as the Cerebras' Wafer-Scale Engine.

\subsection{Tracking the Use of Export-Controlled Chips Through Code Signatures}
Tencent has publicly advertised their use of NVIDIA H100 and A100 GPUs in papers accompanying the release of at least two of their recent models. In May 2024, Tencent released the HunyuanDiT model~\cite{liHunyuanDiTPowerfulMultiResolution2024}, a text-to-image diffusion transformer comparable to OpenAI’s DALL·E 3~\cite{betkerImprovingImageGeneration2023}. The repository for the project explicitly says that experiments for the model were run on NVIDIA A100 GPUs, and of lesser importance, V100s.\footnote{\href{https://github.com/RitwikGupta/HunyuanDiT}{https://github.com/RitwikGupta/HunyuanDiT}. We link to a forked version of the codebase to ensure that it remains frozen in time as analyzed.} In September 2024, Tencent announced the GameGen-O diffusion transformer model that generates open-world video games.\footnote{\href{https://x.com/_akhaliq/status/1834590455226339492}{Tweet by AK}} While the project website no longer exists,\footnote{\href{https://gamegen-o.github.io/}{https://gamegen-o.github.io/}} snapshots of the old website can be found through web archives.\footnote{\href{https://web.archive.org/web/20240000000000*/https://gamegen-o.github.io/}{Wayback Archive copy}} In November of 2024, the same research team published a paper on GameGen-X~\cite{cheGameGenXInteractiveOpenworld2024}, seemingly the same project with a new name, though now without explicit reference to Tencent. The paper details the use of eight NVIDIA H100s for their experiments.

Let us assume that Tencent had not explicitly stated that they are using export-controlled GPUs, can we find signatures in their code that reveal similar details? To a degree, yes. In this section, we reverse engineer candidate GPUs used by Tencent by analyzing representative code signatures in multiple Tencent codebases. We do this iteratively, excluding classes of GPUs at every step.

\begin{figure}[h]
    \centering
    \includegraphics[width=\linewidth]{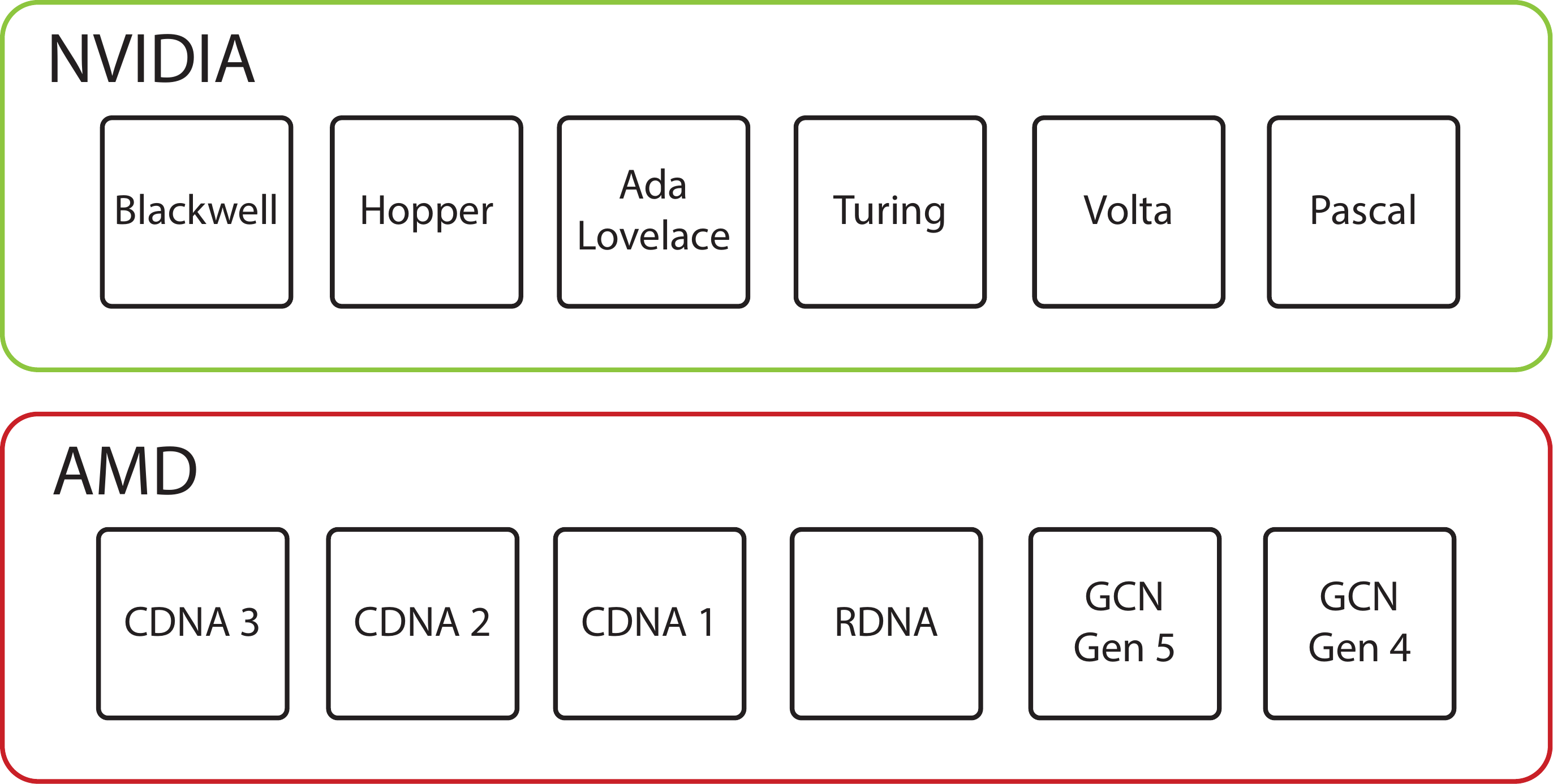}
    \caption{A non-comprehensive list of GPUs under consideration with the potential to train AI models.}
    \label{fig:starting-point}
\end{figure}

\textbf{NVIDIA or AMD?} There are two providers of GPUs that currently matter for machine learning model training: NVIDIA and AMD.\footnote{We apologize to our colleagues at Intel and Apple. There are also a plethora of deep learning accelerator (DLA) manufacturers that are of interest, but we exclude those from consideration due to their lack of market proliferation.} When working with multiple GPUs, efficient communication becomes a bottleneck. The NVIDIA Collective Communications Library (NCCL) provides low-level primitives for multi-GPU and multi-node communications. Notably, NCCL is only for NVIDIA GPUs. It does not work with AMD GPUs.

The training scripts for Hunyuan-Large\footnote{\href{https://github.com/RitwikGupta/Tencent-Hunyuan-Large/blob/main/train/train.sh\#L6}{Link to repository}} and Hunyuan-DiT\footnote{\href{https://github.com/RitwikGupta/HunyuanDiT/blob/main/hydit/train_deepspeed.py\#L180}{Link to repository}} explicitly use the NCCL library for multi-GPU communication demonstrating that they are using NVIDIA GPUs for training.

\begin{figure}[h]
    \centering
    \includegraphics[width=\linewidth]{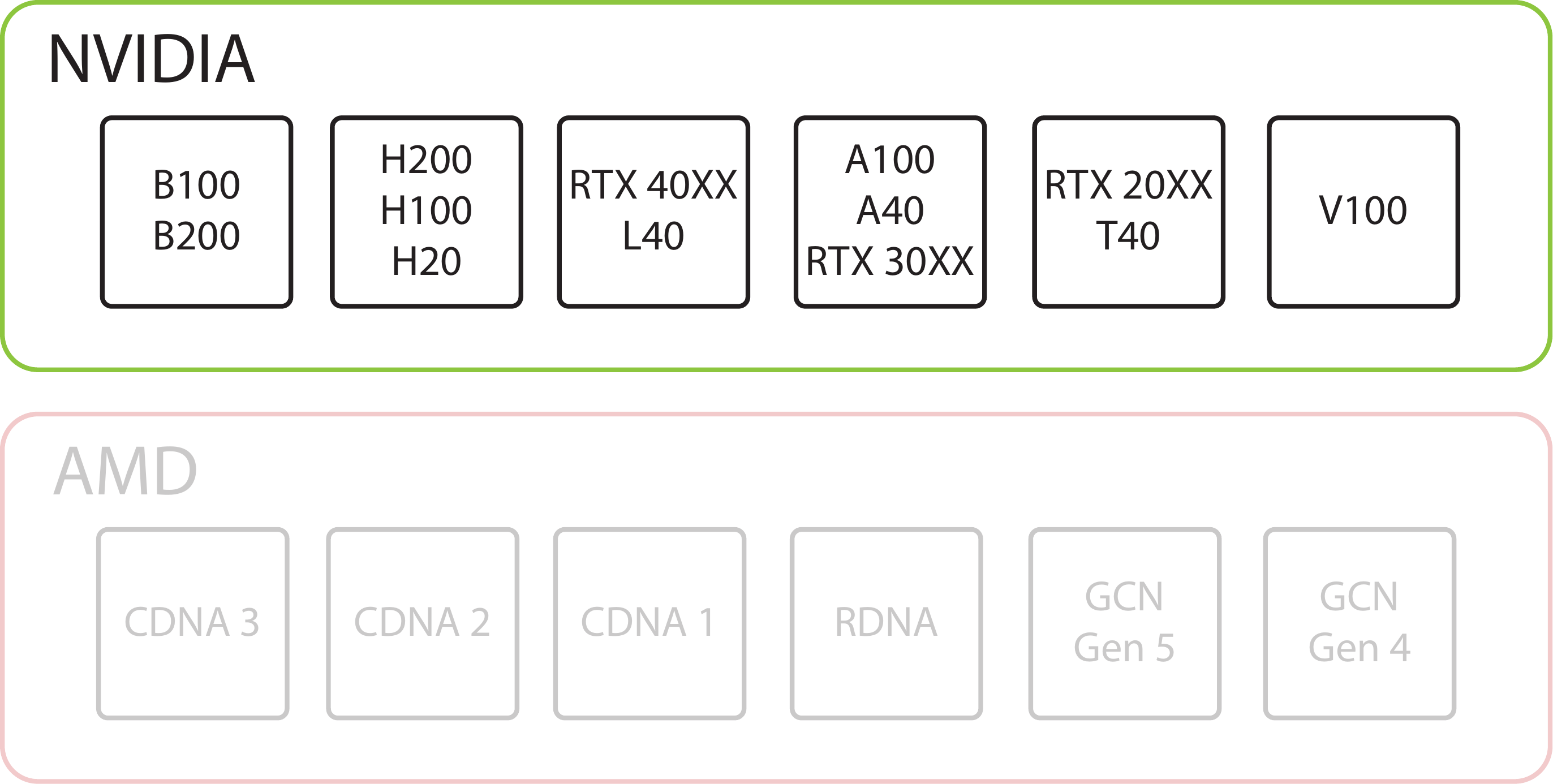}
    \caption{The use of NCCL eliminates AMD GPUs from consideration entirely.}
    \label{fig:nccl}
\end{figure}

\textbf{Utilizing bfloat16.} NVIDIA introduced bfloat16 support with its Ampere micro-architecture which includes the NVIDIA 30XX, 40XX, A100, and other GPUs. Prior to this micro-architecture release, no NVIDIA GPUs supported bfloat16 training.

In the codebases for both Hunyuan-Large\footnote{\href{https://github.com/RitwikGupta/Tencent-Hunyuan-Large/blob/main/train/ds_zero3_offload_no_auto.json\#L11}{Link to repository}} and HunyuanDiT,\footnote{\href{https://github.com/RitwikGupta/HunyuanDiT/blob/main/kohya_ss-hydit/sd-scripts/hunyuan_train.py\#L388}{Link to repository}} bfloat16 support is explicitly specified. This sets a lower bound for potential GPUs utilized to be the Ampere micro-architecture.

\begin{figure}[h]
    \centering
    \includegraphics[width=\linewidth]{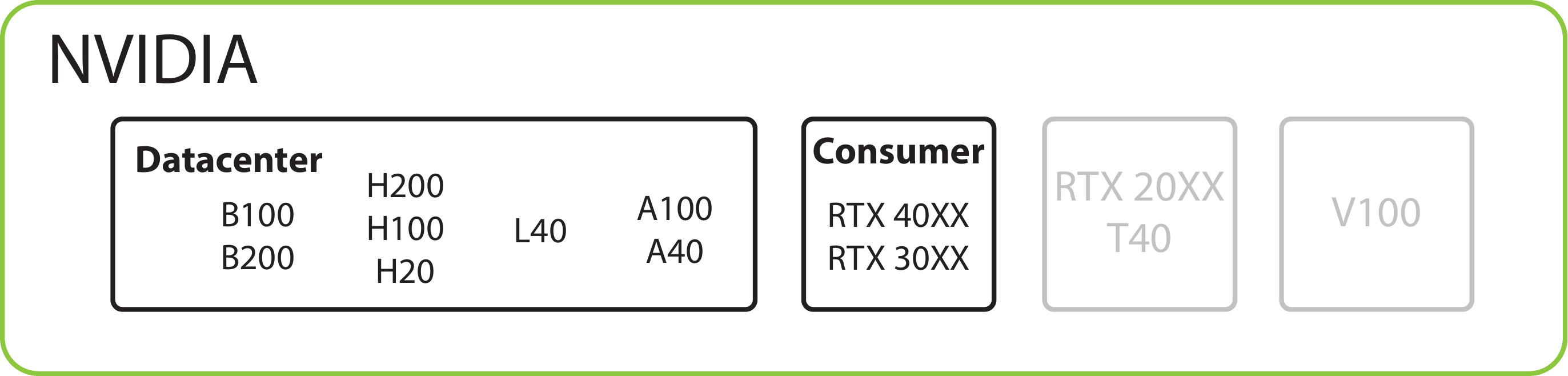}
    \caption{bfloat16 is only available beginning with the Ampere micro-architecture.}
    \label{fig:bfloat16}
\end{figure}

\textbf{Peer-to-peer communications and RDMA.} Peer-to-peer communications via GPUDirect Remote Direct Memory Access (RDMA) drastically improve memory latency during model training. Rather than transfer data from GPU VRAM to CPU RAM and back to another GPU’s VRAM, GPUDirect RDMA allows GPUs to communicate with each other directly via the NVLink or PCIe network fabric locally, or InfiniBand or Ethernet across nodes.

Notably, NVIDIA limits GPUDirect RDMA only to its datacenter GPU offerings. GPUDirect RDMA support for some of the less commonly used GPUs, like the A10 and A40, are not officially documented while the A100 and H100 explicitly support it. As discussed previously, the Hunyuan-Large script configures NCCL to use RDMA over InfiniBand, indicating that the model is trained, minimally, on Ampere datacenter GPUs.

\begin{figure}[h]
    \centering
    \includegraphics[width=\linewidth]{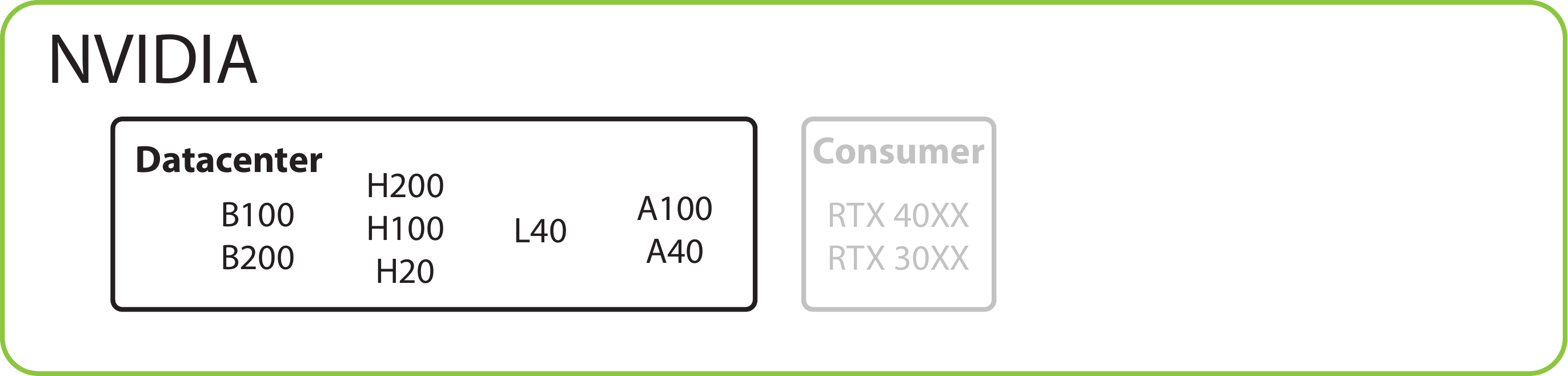}
    \caption{GPUDirect RDMA is a feature that NVIDIA limits to data center GPUs.}
    \label{fig:p2p-rdma}
\end{figure}

We have narrowed our options drastically to a few, modern data center GPUs. This is as far as we can take our analysis based on the code aloneas there are no meaningful feature differences between the Hopper, Ada, and Ampere architectures (besides more cores) to discriminate between them. The Blackwell architecture remains unreleased at the time of writing and largely undocumented. 

\subsection{Avenues of Access}
These public acknowledgments from Tencent corroborate reporting~\cite{yeFocusChinasUnderground2023, linWSJNewsExclusive2023, olcottChineseAIGroups2023} that the PRC is managing to evade U.S. export control. This evasion happens in a variety of different ways. By cross-reference reporting with the publicly available papers, training codebase, and weights from leading PRC AI labs, we not only begin to understand the ways in which PRC companies are accessing restricted chips, but also which models and companies are using which chips, as well well as the techniques being used to overcome having access to chips of lower quality.

\textbf{Stockpiling chips in advance of restrictions.} Advanced GPUs such as the NVIDIA A100 were sold in the Chinese market as there were previously no restrictions on the sale of such components. When it became clear that their access to GPUs would be limited, Chinese companies intensified efforts to stockpile chips.  This practice of stockpiling continues whenever new restrictions are anticipated but not yet implemented~\cite{yangExclusiveChineseFirms2024, shilovChinasChipImports2024}. Aforementioned discussions concerning the future regulation of the H20 may lead to further stockpiling in the months ahead.

\textbf{Illegally exporting and acquiring physical chips.} PRC companies on the entity list also have options when it comes to procuring restricted chips, namely purchasing from vendors and illegal markets~\cite{yeFocusChinasUnderground2023} or using subsidiaries~\cite{linWSJNewsExclusive2023} that are not on the entities list. Per \textit{Reuters} reporting in 2023~\cite{yeFocusChinasUnderground2023}, third-party sellers in China, who range from individuals to entire marketplaces of restricted electronics, typically come by their wares by either buying up excess stock on the black market or by using a company outside China to initially acquire the chips without limitations.

Both of these approaches allow PRC companies to gain physical access (though in limited amounts) to chips including the A100 and the H100. As detailed in \Cref{sec:bypass}, the precise configuration of networking parameters, such as the InfiniBand timeouts and traffic class, provides a clue that Tencent is utilizing on-premise GPUs where such fine optimizations are possible. Cloud computing environments rarely afford the ability to tune networking parameters such as these.

\textbf{Renting access or otherwise remotely accessing chips.} Beyond direct acquisition, companies in the PRC are also able to access limited chips through cloud computing companies. The \textit{Financial Times} identified two companies currently sanctioned by the US, iFlytek and SenseTime, who were using AI-Galaxy, a cloud computing company, to access A100 chips~\cite{olcottChineseAIGroups2023}.

It is worth noting that NVIDIA H20s are not available for rent through cloud services. No U.S.-based cloud providers sell H20 GPU hours, and neither do Baidu Cloud, Alibaba Cloud, Huawei Cloud, or Tencent Cloud.\footnote{Based on our review of the offerings of Baidu Cloud, Alibaba Cloud, Amazon Web Services, Google Cloud Platform, Microsoft Azure, Runpod, Lambda Cloud, Vast AI, TensorDock, and OVH Cloud.}

\section{A Reckoning for U.S. Semiconductor Export Controls}

Advances in machine learning have eroded the moat that the United States initially built to limit access to the latest GPUs via its existing export controls. Moving forward, Washington can either choose to double down on the idea of export controls, limiting more and more types of hardware, or it can recognize the inefficacy of broad semiconductor export controls and pivot to a more nuanced approach.

\textbf{The evidence presented in this paper suggests that export controls on AI are less effective than Washington might like, and that semiconductor export controls appear to be leaky proxies.} So, what are the alternatives?

At present, the United States is attempting to limit the military use of AI software without placing restrictions on the export of AI software itself. It is unlikely that the United States should, or would, control the spread of AI software. For one, this path is fraught with technical and legal risks. AI is inherently dual-use; models built to detect illegal fishing vessels~\cite{paoloXView3SARDetectingDark2022} can equivalently be used to identify military ones. Controlling the latter limits the utility of the former. Further, the Ninth Circuit established the protection of code as First Amendment-protected speech in \textit{Bernstein v. United States}. AI is also ``code'' and any attempts to control it could be met with similar legal rebuke.

Alternatively, the United States might instead establish a robust verification regime by first clearly identifying use cases that harm public safety~\cite{shokerNewToolsAre2024}, then establishing clear, continually updated benchmarks\footnote{This can be done periodically as with \href{https://mlcommons.org/benchmarks/training/}{MLPerf}.} to establish a frontier for those use cases, and finally comparing the frontier to the abilities of the Chinese domestic ecosystem. The U.S. government---likely the Department of Commerce via the newly-established NIST AI Safety Institute---might establish competitive, and continuously updated, benchmarks for those tasks. We can then train models that attempt to address these tasks on constrained hardware, giving a clear understanding of where the foreign frontier may be.

\textbf{Some will argue for expanded export controls, including the removal of hardware loopholes that provide useful chips to the PRC.} As the diminishing returns of current export controls become increasingly apparent, policymakers might be tempted to double down on export controls, removing loopholes and expanding their scope.

Certain consumer GPUs such as the RTX 4090 are restricted for export to China due to exceeding the performance thresholds\footnote{```Total processing performance' (`TPP') is 2 $\times$ `MacTOPS' $\times$ `bit length of the operation', aggregated over all processing units on the integrated circuit'' as per ECCN 3A090. 330.3 MacTOPs x 16 bit length = 5,284.8 TPP, using Peak FP16 Tensor TFLOPS with FP16 Accumulate from the \href{https://images.nvidia.com/aem-dam/Solutions/Data-Center/l4/nvidia-ada-gpu-architecture-whitepaper-v2.1.pdf}{Ada GPU Architecture Whitepaper}.} outlined in the regulations published in October 2023. NVIDIA created the RTX 4090D to comply with the regulations and has made it available for sale in China.

An expansion of controls, or even closing of loopholes, will have to reckon with powerful financial incentives at play. GPU manufacturers will continue to be lured by the lucrative Chinese video game hardware market---valued at RMB\textyen240.3 billion (\$33.31 billion) at the end of 2023~\cite{statistaGamingHardwareChina}---and attempt to fill the market by producing marginally sub-threshold alternatives. Chinese AI companies will then use those same sub-threshold GPUs to skirt export controls, continuing the cycle described above. As demonstrated in this work, these ``weaker'' chips can still be effectively utilized to train state-of-the-art models.

The U.S. government may also be tempted to lower TPP limits to be inclusive of all chips, even if it comes at huge financial costs to U.S. companies. However, even if those financial costs were acceptable, the U.S. government would continue to face the same challenges that it does today. 

\textbf{Further restricting PRC access to high-end chips will be insufficient in limiting their AI gains.} As highlighted in this paper, the PRC has numerous avenues for accessing export-controlled chips, including stockpiling chips presently available, in anticipation of future restrictions. Furthermore, the PRC’s domestic chip production, such as of Huawei's Ascend 910-series GPUs, is accelerating~\cite{liangChinaUnveilsDomestic2024}. Despite facing issues with yield and cost~\cite{nasirChineseThinkTank2024}, it is believed that domestic hardware quality could, in due course, match the performance of U.S. GPUs~\cite{morganHuaweisHiSiliconCan2024}. The biggest hurdle, we argue, will be China's ability to create adequate software that exploits their hardware properly~\cite{mcmorrowHuaweisBugriddenSoftware2024}.

\section{Conclusion}
Regardless of the approach it takes, the United States must closely follow not only capability gains in Chinese models but also gains in software-hardware efficiency. Close monitoring and evaluation of publicly available papers, codebases, and weights from leading PRC AI labs, exemplified in this paper, can provide insights into not only the use of export-controlled chips by Chinese entities, but also the techniques, tools, and libraries that labs use to make the most of non-export controlled chips. This continuous monitoring may also allow for the detection of potential increased use of domestically produced chips and the capture of useful information concerning broader trends in Chinese machine learning. 

\section*{Acknowledgements}
Authors, as part of their affiliation with UC Berkeley, were supported in part by the National Science Foundation, U.S. Department of Defense, Founders Pledge Fund, Ford Foundation, and/or the Berkeley Artificial Intelligence Research (BAIR) industrial alliance program. We thank Giscard Biamby, Nick Garcia, Rodolfo Corona, and other friends and collaborators who reviewed this work for technical correctness and rigor prior to publication.


\clearpage
\bibliographystyle{plain}
\bibliography{reference}


\end{document}